\documentstyle[12pt]{article}
\title{
FERMIONS ON HALF-QUANTUM VORTEX}
\author{ G.E. Volovik\\
Low Temperature Laboratory\\
Helsinki University of Technology\\
Otakaari 3A, 02150 Espoo, Finland\\
and\\
L.D. Landau Institute for Theoretical Physics, \\
Kosygin Str. 2, 117940 Moscow, Russia\\
}
\begin{document}
\maketitle
\begin{abstract}
{The spectrum of the fermion zero modes in the vicinity of the  vortex with
fractional winding number is discussed. This is inspired by the observation of
the 1/2 vortex in high-temperature superconductors \cite{Kirtley1996}. The
fractional value of the winding number leads to the fractional value of the
invariant, which describes the topology of the energy spectrum of fermions.
This results in the phenomenon of the "half-crossing": the spectrum approaches
zero but does not cross it, being captured at the zero energy level. The
similarity with  the phenomenon of the fermion  condensation is discussed. }
\end{abstract}

\vfill \eject

\section{Introduction}

Recently a new topological object in high-temperature superconductor -- the
vortex with the fractional winding number $N=1/2$ -- has been observed
\cite{Kirtley1996}. This 1/2-vortex was predicted  in
Ref.\cite{Geshkenbein1987}.

As a rule, the fractional vortices are coupled to the topological or
nontopological surfaces: In $^3$He-A  where such
vortices should exist \cite{VolovikMineev1976} but still have not been
identified, they represent  the termination lines of the topological
$Z_2$-solitons. The same occurs for the 1/2 vortices   discussed in heavy
fermionic systems \cite{Zhitomirsky1995}. In  high-temperature superconductors
the observed 1/2 vortex is attached to the tricrystal line, which is the
junction of three grain boundaries \cite{Geshkenbein1987,Kirtley1996}. In
$^3$He-B a pair of 1/2 vortices, connected by the nontopological wall, forms a
nonaxisymmetric $N=1$
vortex \cite{Thuneberg1986,SalomaaVolovik1987,Volovik1990}, the latter
vortex has
been experimentally identified in Ref.\cite{Kondo1991}. In
Ref.\cite{Kharadze1995} even the 1/4 quantum vortices were predicted to exist in
$^3$He-A$_2$, they are connected by the nontopological walls discussed in
\cite{SalomaaVolovik1989}.

The $N=1/2$ vortex observed in high-temperature superconductor
carries 1/2 of the magnetic flux quantum $\Phi_0=h/2e$  \cite{Kirtley1996}. In
principle the flux number does not necessarily coincide with the winding
number $N$, thus the objects with the fractional flux below $\Phi_0/2$ are also
possible \cite{VolovikGorkov1984}. They can arise if the time inversion symmetry
is broken \cite{Sigrist1989,Sigrist1995}.

One may expect different interesting phenomena related to the behavior
of the quasiparticles (fermions and bosons) in the presence of fractional
vortices. These vortices are the counterpart of  Alice strings, which appear
in some models of particle physics: the particle travelling around some type of
the  Alice string changes its electric charge to the opposite
\cite{Schwarz1982}.  In  $^3$He-A, the analogous effect is the reversal of the
spin of the quasiparticle upon circling the 1/2 vortex. This behavior results
also in the peculiar Aharonov-Bohm effect, which has been discussed for  1/2
vortices in $^3$He-A \cite{Khazan1985,SalomaaVolovik1987} and  has been modified
for the cosmic Alice strings in \cite{March-Russel1992,Davis1994}.

Here we discuss the behavior of the fermion zero modes -- Caroli-de
Gennes-Matricon (CGM) bound states \cite{Caroli1964} in the core of
vortices. As was found in Ref.\cite{Q-modes-Index} there is some kind of the
index theorem for the CGM levels. It states that the number of  branches of the
quasiparticle spectrum, which as function of the angular momentum $Q$ cross
the zero energy, equals the winding number $N$ per each of two spin projections.
For the conventional $N=1$ vortex in $s$-wave superconductors this results in
two branches of fermion zero modes corresponding to two spin projections.  The
problem is what happens if the winding number per spin is fractional.  This is
the relevant case for the
$N=1/2$ vortices in heavy fermionic and high-T superconductors and for the
$N=1/4$  vortices in
$^3$He-A$_2$.  (This is not related to the 1/2 vortex in
$^3$He-A, because $^3$He-A can be effectively represented as an equal mixture
of spin-up and spin-down components, and the 1/2 vortex corresponds to the $N=1$
vortex in one component  with $N=0$ for another component. Thus the winding
number per spin appears to be integer for each component.)

Here we shall use the model of 1/2 vortex discussed in \cite{Zhitomirsky1995}.
The result is schematically presented in Fig.1. While for the
$N=1$ vortex the spectrum $E(Q)$ crosses zero (Fig.1a), for the $N=1/2$ vortex
the spectrum instead of crossing asymptotically approaches zero (Fig.1b). The
Fig.1c shows the spectrum for the case when two 1/2 vortices together form the
$N=1$ vortex.

\section{"Half-crossing".}

Let us consider the unconventional superconductivity of the $E_{1g}$
representation, which corresponds to the $d$-wave pairing.  The symmetry class
which gives rise to the 1/2 vortex is characterized by the following gap
function:
$$
\Delta ({\bf k},{\bf r})=\Delta_0({\bf r})~ e^{i\Phi({\bf
r})}~\hat k_z ~\hat{\bf k}\cdot\hat {\bf d}({\bf r})
~~,
\eqno(1)$$
where $\hat{\bf k}$ is the unit vector along the linear momentum ${\bf k}$ of
the quasiparticle; the  unit real vector  $\hat {\bf d}$ of the spontaneous
anisotropy, the condensate phase $\Phi$ and the gap amplitude
$\Delta_0$ are coordinated dependent in  the presence of the 1/2 vortex. We
assume the simplest form for the 1/2 vortex with
$$\Delta_0({\bf r})= \Delta_0(r)~~,~~ \Phi({\bf r})={\phi\over 2} ~~,~~\hat
{\bf d}({\bf r})=\hat x\cos{\phi\over 2} + \hat y\sin{\phi\over 2}
~~,
\eqno(2)$$
where $z,r,\phi$ are the cylindrical coordinates. The changes of the sign of
the gap function due to the $\pi$ winding of the phase $\Phi=\phi/2$  is
compensated by the change of sign of the  $\hat {\bf d}$ field.

The fermionic
spectrum is obtained from the Bogoliubov Hamiltonian, which for the  given spin
projection is
$2\times 2$ matrix
$$
{\cal H}=\hat \tau_3 \varepsilon({\bf p})
+\hat \tau_1 Re\Delta ({\bf p},{\bf r})
-\hat \tau_2 Im\Delta ({\bf
p},{\bf r})~~,
\eqno(3)$$
where $\hat \tau_a$ are the Pauli matrices, ${\bf p}=-i\vec\nabla$,
$\varepsilon({\bf p})=(p^2-p_F^2)/2m$.

In quasiclassical approximation ($p_F\xi \gg 1$ where $\xi$ is the core size)
the relevant variables are the transverse momentum ${\bf q}$ of the incident
quasipartice with $q^2=p_F^2-p_z^2 $ and its impact parameter $b$. The angular
momentum of the quasipartice is $Q=qb$. For the Ansatz (2) there is an important
difference between positive and negative values of the impact parameter $b$, see
Fig.2. For one sign of $b$ the quasiparticle trajectory always crosses the
surface at which the classical energy takes zero value: $E({\bf k},{\bf
r})=\sqrt{\varepsilon^2({\bf k}) +\vert \Delta ({\bf k},{\bf r})\vert^2}=0$,
while for the opposite sign of $b$ the energy is always nonzero. This leads to
the asymmetry of the fermionic spectrum in Fig.1b.

Let us consider the simplest example of ${\bf q}=q\hat y$. In this case the
surface of the zero value of the gap in Eq.(1) is the half-plane $y=z=0$,
$x>0$ (see Fig.2). The Hamiltonian  is
$$ {\cal H}= -i \hat \tau_3 {q\over m}  \nabla_y  + {q k_z\over 2
p_F^2}\Delta_0(r)~[\hat \tau_1
\sin\phi -\hat \tau_2
(1-\cos\phi)]~~.
\eqno(4)$$
If $b>0$ and large enough, then in the vicinity of the cross point of the
trajectory with the zero-energy surface  one has
$\phi\approx y/b \ll 1 $ and the third term can be considered as small
perturbation. The first two terms have the normalizable zero-energy solution,
which for $b\gg\xi$ has the form
$$
\Psi^{(0)}(y)\propto
[1- \hat \tau_2{\rm sign}(k_z b)]~~
e^{- {m \vert k_z \vert\Delta_0(\infty)\over
4b p_F^2}    y^2 }                   ~~.\eqno(5)$$
The third term averaged over this wave function gives the following energy
spectrum:
$$E(q,b)\sim {q\over mb}  ~~,~~b>0 ~~.\eqno(6)$$
Since the effective $q$ is on order  of $p_F$, the spectrum in terms of the
angular momentum $Q=qb$ has the following form
$$E_{{\bf q}=q\hat y}(Q)\sim {E_F \over Q} ~~,~~b>0~~, ~~b\gg \xi
~~,\eqno(7)$$
where $E_F=p_F^2/2m$ is the Fermi-energy.  At large $b$ this
spectrum approaches the zero energy level: these asymptotic midgap states far
from the half-quantum vortex reflect the property of the fermionic bound states
in the $\pi$-soliton or in the Josephson junction  with the phase difference
$\pi$, which emanates from the 1/2 vortex.

If $b<0$ the trajectory does not cross the zero value of the classical energy
and the behavior of the spectrum follows that for the conventional vortex. This
can be seen on the model example of the large core size compared with $\xi$. In
this case the Hamiltonian in Eq.(4) can be expanded near the point $r=\vert
b\vert$, $\phi=0$:
$$ {\cal H}= -i \hat \tau_3 {q\over m}  \nabla_y  + {q k_z\over 2
p_F^2}\Delta_0(b)~(\hat \tau_1
{y\over \vert b\vert}- 2 \hat \tau_2 )~~.
\eqno(8)$$
This gives the energy spectrum
$$E(q,b)= {q \vert k_z\vert\over
p_F^2}\Delta_0(b)~~,~~b<0 ~~, \eqno(9)$$
and for small $b$, when $\Delta_0(b)=\vert b\vert \Delta'$ the ordinary linear
dependence on the angular momentum is obtained
$$E_{{\bf q}=q\hat y}(Q)\sim - Q~{\Delta_0^2 \over E_F}  ~~,~~b<0~~, ~~\vert b
\vert < \xi   ~~.\eqno(10)$$

Together with the Eq.(7) this gives the spectrum in Fig.1b.
For the opposite direction of the transverse momentum ${\bf q}=-q\hat y$ the
spectrum is inverted, see the right part of Fig.1c. Thus when two $N=1/2$
vortices form the $N=1$ vortex the topology of the spectrum in Fig.1a is
restored: spectra in Fig.1a and Fig.1c have the same topology, which corresponds
to one branch crossing zero. If one considers the other orientations of  ${\bf
q}$ one obtains either the spectrum in Fig.1b or the inverted one. Thus for
each  ${\bf q}$ the energy spectrum $E_{\bf q}(Q)$ in terms of the angular
momentum $Q$ "half-crosses" zero.

\section{Relation to the index theorem.}

The "half-crossing" is also supported by the
index theorem and corresponds to the fractional value of the topological
invariant introduced in \cite{Q-modes-Index}.  The number of  the
anomalous branches of the spectrum $E_{\bf q}(Q)$, which at given ${\bf q}$
cross zero as a fuction of $ Q$, coincides with the number of
 the point zeroes of the Bogoliubov Hamiltonian in its classical limit. For
${\bf q}=q\hat y$ the classical expression follows from Eq.(4)
$$ {\cal H}_{\rm class}(p_y,y,x)=  \hat \tau_3 {q\over m}  p_y  + {q k_z\over 2
p_F^2}\Delta_0(r)~[\hat \tau_1
\sin\phi -\hat \tau_2
(1-\cos\phi)]~~,
\eqno(11)$$
which can be written in terms of
the unit vector  $\hat m(\vec s)$ in the 3-d space of parameters $\vec
s=(p_y,x,y)$:
$${\cal H}_{\rm class}(\vec s)=\vec \tau\cdot \vec m(\vec s)~~.
\eqno(12)$$
The components of $\hat m(\vec s)$ are
$$E\hat m_3(\vec s)=
qp_y/m ~~,~~E\hat m_1(\vec s)={q k_z\over 2
p_F^2}{\Delta_0(r)\over r}~y ~~,~~E\hat m_2(\vec s)=-{q k_z\over 2
p_F^2}{\Delta_0(r)\over r}~(r-x)
~~,\eqno(12)$$
where $E$ is the classical energy:  $E^2(\vec s)={\cal H}^2_{\rm class}(\vec
s)$. The $\hat m(\vec s)$ vector field, shown schematically in Fig.3,
corresponds to the hedgehog with the $\pi_2$ topological charge 1/2:
$$N_{\rm zm}={1 \over {8\pi }}\int_{\sigma } dS^ie_{ikl}\Bigl(\hat m \cdot { {
\partial \hat m} \over {\partial s_k}} \times { { \partial \hat m } \over {
\partial s_l } } \, \Bigr) \hskip2mm ,
\eqno(13)$$
where the integral is over the hemisphere  $\sigma$ about
the hedgehhog. This  fractional value of the topological invariant $N_{\rm
zm}=1/2$, which  characterizes  the 1/2 vortex, is responsible for the
"half-crossing" in correspondence with the index theorem.

\section{Discussion.}

The spectrum of the fermion zero modes in the vicinity of the 1/2 vortices
has peculiar topology shown in Fig.1b-c. This follows from the fractional value
of the topological invariant, which describes the crossing of the zero energy.
This topological invariant coincides with the winding
number $N$ of the vortex. For vortices with integer $N$, there are
$N$ branches (per spin) which cross zero. For the $N=1/2$ vortex, this invariant
is fractional and this corresponds to the "half-crossing" in Fig.1b: the
spectrum approaches zero but does not cross it, being captured at the zero
energy level. Thus far from the half-quantum vortex the
spectrum approaches the manifold of the midgap states with zero energy, which is
the property of the fermionic bound states in the surface which emanates
from the
1/2 vortex. This surface can be the $\pi$-soliton (in the heavy fermionic
superconductors) or the Josephson junction  with the phase difference $\pi$ (in
the high-T superconductors). The existence of the midgap manifold is the robust
property of such surfaces \cite{MakhlinVolovik1995} (see also
\cite{YangHu1994}), in the absence of impurities and if the time inversion
symmetry is not violated.

The situation with the zero-energy manifold is similar to the
phenomenon of the fermionic condensate discussed in \cite{Khodel1990,Khodel1994}
(the extended Van Hove singularity observed in oxide superconductors
\cite{Yokoya} can be the manifestation of this phenomenon, which is favourable
just in the vicinity of the saddle point of spectrum \cite{Volovik1994}). It is
interesting that the existence of the  fermionic condensate is also related to
the splitting of the
$N=1$ vortex into two $N=1/2$ vortices but in the momentum ${\bf k}$ space
\cite{Volovik1991}. The $N=1$ vortex  corresponds to the Fermi-surface in the
momentum space. Its splitting results in the Fermi-band with zero energy, which
appears between the $N=1/2$ vortices in the momentum space. In both cases the
Fermi-condensate is characterized by a large density of states.

This work was supported through the ROTA co-operation plan of the Finnish
Academy and the Russian Academy of Sciences and by the Russian Foundation for
Fundamental Sciences.

\begin{figure}[h]
\begin{center}
\caption[Fig.1]{%
Topology of the fermion zero modes in vortices with winding number $N$. (a)
In $N=1$ vortex there is one branch of the spectrum, which as a function of the
angular momentum $Q$ crosses zero energy.  (b)  For the vortex with
fractional winding number $N=1/2$ the spectrum approaches zero but does
not cross it, being captured at the zero energy level.
(c) The topology of the spectrum in the case of a pair of $N=1/2$ vortices,
which forms the $N=1$ vortex, is the same as in Fig.1a.  }
\label{Fig.1}
\end{center}
\end{figure}

\begin{figure}[h]
\begin{center}
\caption[Fig.2]{%
Demonstration of the origin of the spectral asymmetry in the $N=1/2$ vortex. If
the impact parameter $b$ is positive  the quasiparticle trajectory always
crosses the surface at which the classical energy takes zero value. For the
opposite sign of $b$ the classical energy is always nonzero.
}
\label{Fig.2}
\end{center}
\end{figure}

\begin{figure}[h]
\begin{center}
\caption[Fig.3]{%
Index theorem for the $N=1/2$ vortex. In the 3D space of the parameters $\vec
s=(p_y,y,x)$ there is a line of zeroes of the classical energy, which terminates
on the monopole in the field of the unit vector $\hat m$,
which characterizes the Hamiltonian. The $\pi_2$ topological charge of the
monopole is fractional and equals
$N=1/2$. This is the topological origin of the "half-crossing" of the spectrum
in Fig.1b. }
\label{Fig.3}
\end{center}
\end{figure}

\end{document}